\documentclass[12pt]{nature}
\usepackage{graphicx}
\usepackage{times}
\usepackage{booktabs} % Nice tables
\usepackage{morefloats}
\usepackage[hyphenbreaks]{breakurl}

\def\apgt{\ {\raise-.5ex\hbox{$\buildrel>\over\sim$}}\ }
\def\aplt{\ {\raise-.5ex\hbox{$\buildrel<\over\sim$}}\ }
\def\lteq{\ {\raise-.5ex\hbox{$\buildrel<\over-$}}\ }

\def\aap{\ {A\&A}\ }

\def\apjs{\ {ApJS}\ }

\def\nat{\ {Nat}\ }

\spacing{1}

\title{The Ecological Impact of High-performance Computing in
Astrophysics}

\author{
Simon Portegies Zwart
}

\begin{document}
\date{}

%\pagerange{ -- } \pubyear{2011} 

\maketitle

%Mdisk = 0.1Msun consisent with 
%https://www.cfa.harvard.edu/news/su201542

\begin{affiliations}
\item Leiden Observatory, Leiden University, PO Box 9513, 2300 RA,
Leiden, The Netherlands 
\footnote{Non-anonymous Dutch scientists.}
\end{affiliations}

%%2004come.book..153D,

%%
\begin{abstract}

The importance of computing in astronomy continues to increase, and so
is its impact on the environment. When analyzing data or performing
simulations, most researchers raise concerns about the time to reach a
solution rather than its impact on the environment. Luckily, a
reduced time-to-solution due to faster hardware or optimizations in
the software generally also leads to a smaller carbon footprint. This
is not the case when the reduced wall-clock time is achieved by
overclocking the processor, or when using supercomputers.

The increase in the popularity of interpreted scripting languages, and
the general availability of high-performance workstations form a
considerable threat to the environment. A similar concern can be
raised about the trend of running single-core instead of adopting
efficient many-core programming paradigms.

In astronomy, computing is among the top producers of green-house
gasses, surpassing telescope operations. Here I hope to raise the
awareness of the environmental impact of running non-optimized code on
overpowered computer hardware.

\end{abstract}

\section{Carbon footprint of computing}

The fourth pillar of science, simulation and modeling, already had a
solid foothold in 4th-century astronomy
\cite{2016Sci...351..482O,2006Natur.444..587F}, but this discipline
flourished with the introduction of digital computers. One of its
challenges is the carbon emission caused by this increased
popularity. Unrecognized as of yet by UNESCO \cite{EOLSS2020} the
carbon footprint of computing in astrophysics should be emphasized.
One purpose of this document is to raise this awareness.

\begin{figure}
\includegraphics[width=\columnwidth]{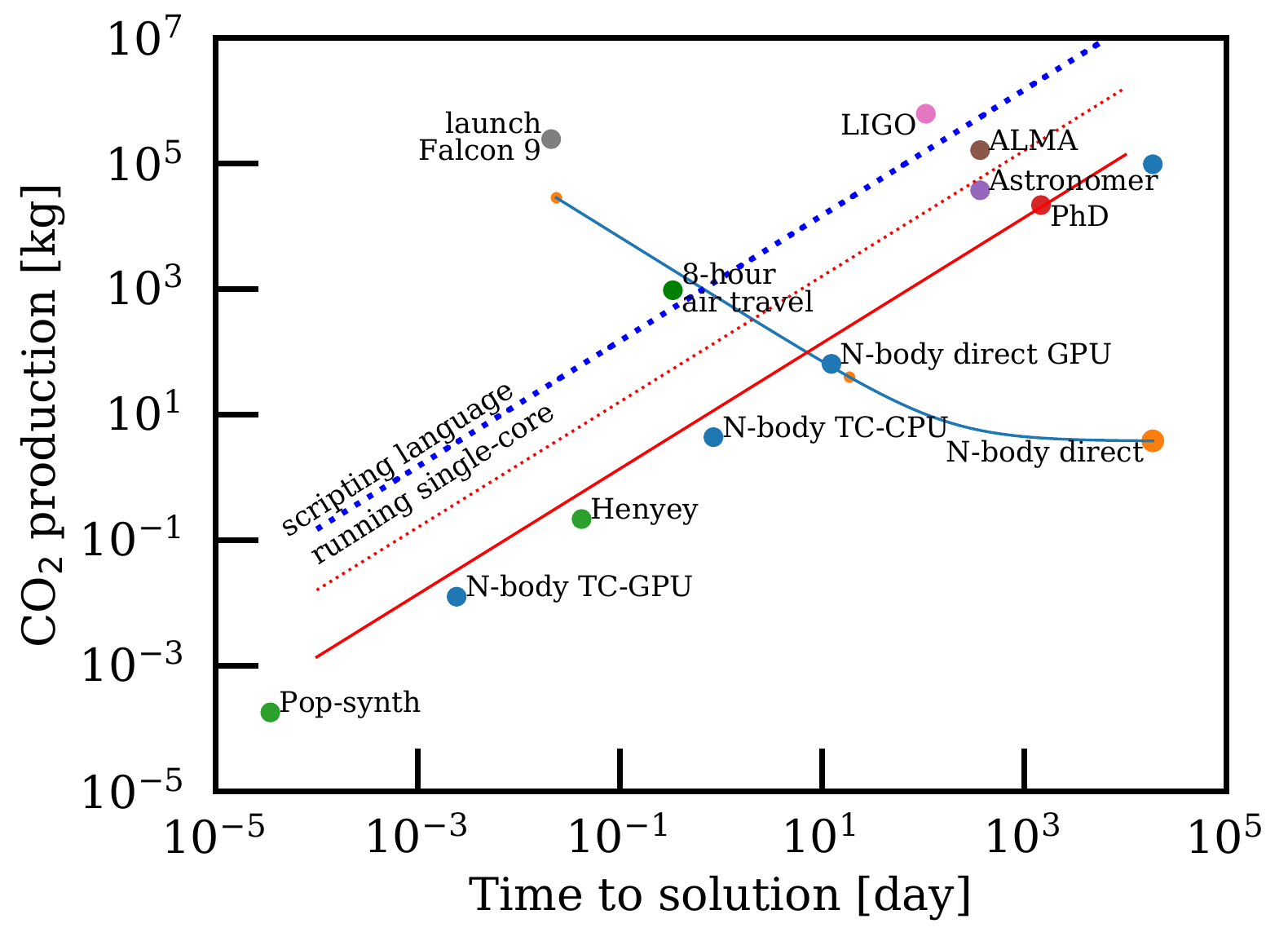}
\caption{\bf CO$^{2}$ emission (in kg) as a function of the time to
solution (in days) for a variety of popular computational techniques
employed in astrophysics, and other activities common among
astronomers \cite{ACHTEN2013352,2019arXiv191205834S}. The solid red
curve gives the current individual world-average production, whereas
the dotted curves give the maximum country average. The LIGO carbon
production is taken over its first 106-day run (using $\sim
180$\,kW) \cite{LIGO_Design}, and for ALMA a 1-year average
\cite{2014era..conf30201D}. A Falcon~9 launch lasts about 32 minutes
during which $\sim 110\,000$\,liters of highly refined kerosene is
burned.
%%%%
The tree-code running on GPU is performed using $N=2^{20}$
particles. The direct N-body code on CPU (right-most blue bullet)
was run with $N=2^{13}$ \cite{2007NewA...12..641P}, and the other
codes with $N=2^{16}$. All performance results were scaled to
$N=2^{20}$ particles. The calculations were performed for 10 N-body
time units \cite{1986LNP...267..233H}. The energy consumption was
computed using the scaling relations of
\cite{DBLP:journals/corr/abs-1304-7664} and a conversion of KWh to
Co$_2$ of $0.283$\,kWh/kg. The blue dotted curve shows the estimated carbon emission when these calculations would have been implemented
in Python running on a single core. The solid blue curve to the
right, starting with the orange bullet shows how the performance and
carbon production changes while increasing the number of compute
cores from 1 to $10^6$ (out of a total of $7\,299\,072$, left-most
orange point) using the performance model by
\cite{heinrich:hal-01523608}. }
\label{fig:tts_CO2_emission}
\end{figure}

In figure\,\ref{fig:tts_CO2_emission}, we compare the average Human
production of CO$_2$ (red lines) with other activities, such as
telescope operation, the emission of an average astronomer
\cite{2019arXiv191205834S} and finishing a (four year) PhD
\cite{ACHTEN2013352}.

While large observing facilities cut down on carbon footprint by
offering remote operation, the increased speed of computing resources
can hardly be compensated by their increased efficiency. This also is
demonstrated in figure\,\ref{fig:tts_CO2_emission}, where we compare
measurements for several popular computing activities. These
measurements are generated using the Astrophysical Multiuser Software
Environment \cite{2018araa.book.....P}, in which the vast majority of
the work is done in optimized and compiled code.

We include simulations of the Sun's evolution from birth to the
asymptotic giant branch using a Henyey solver
\cite{2011ApJS..192....3P} and parametrized population-synthesis
\cite{1996A&A...309..179P} (green bullets).

We also present timings for simulating the evolution of a
self-gravitating system of a million equal-mass point-particles in a
virialized Plummer sphere for 10 dynamical time-scales. These
calculations are performed by direct integration (with the 4th-order
Hermite algorithm) and using a hierarchical tree-code (with leapfrog
algorithm). Both calculations are performed on CPU as well as with
graphics processing unit (GPU). Not surprisingly, the tree-code
running single GPU is about a million times faster than the
direct-force calculations on CPU; One factor $1000$ originates from
the many-cores of the GPU \cite{2010ProCS...1.1119G}, and the other
from the favorite scaling of the tree algorithm
\cite{1986Natur.324..446B}. The trend in carbon production is also
not surprising; shorter runtime leads to less carbon. The emission of
carbon while running a workstation is comparable to the world's
per-capita average.

Now consider single-core versus multi-core performance of the direct
$N$-body code in figure\,\ref{fig:tts_CO2_emission}. The blue bullet
to the right gives the single-core workstation performance, but the
large orange bullet below it shows the single-core performance on
today's largest supercomputer \cite{Green500}. The blue curve gives
the multi-core scaling up to $10^6$ cores (left-most orange point).
The relation between the time-to-solution and the carbon footprint of
the calculations is not linear. When running a single core, the
supercomputer produces less carbon than a workstation (we assumed the
supercomputer to be used to capacity by other users). Adopting
more cores result in better performance, at the cost of producing
more carbon. Similar performance as a single GPU is reached when
running $1000$ cores, but when the number of cores is further
increased, the performance continues to grow at an enormous cost in
carbon production. When running a million cores, the emission of
running a supercomputer by far exceeds air travel and approaches the
carbon footprint of launching a rocket into space.

\section{Concurrency for lower emission}

\begin{figure}
\includegraphics[width=\columnwidth]{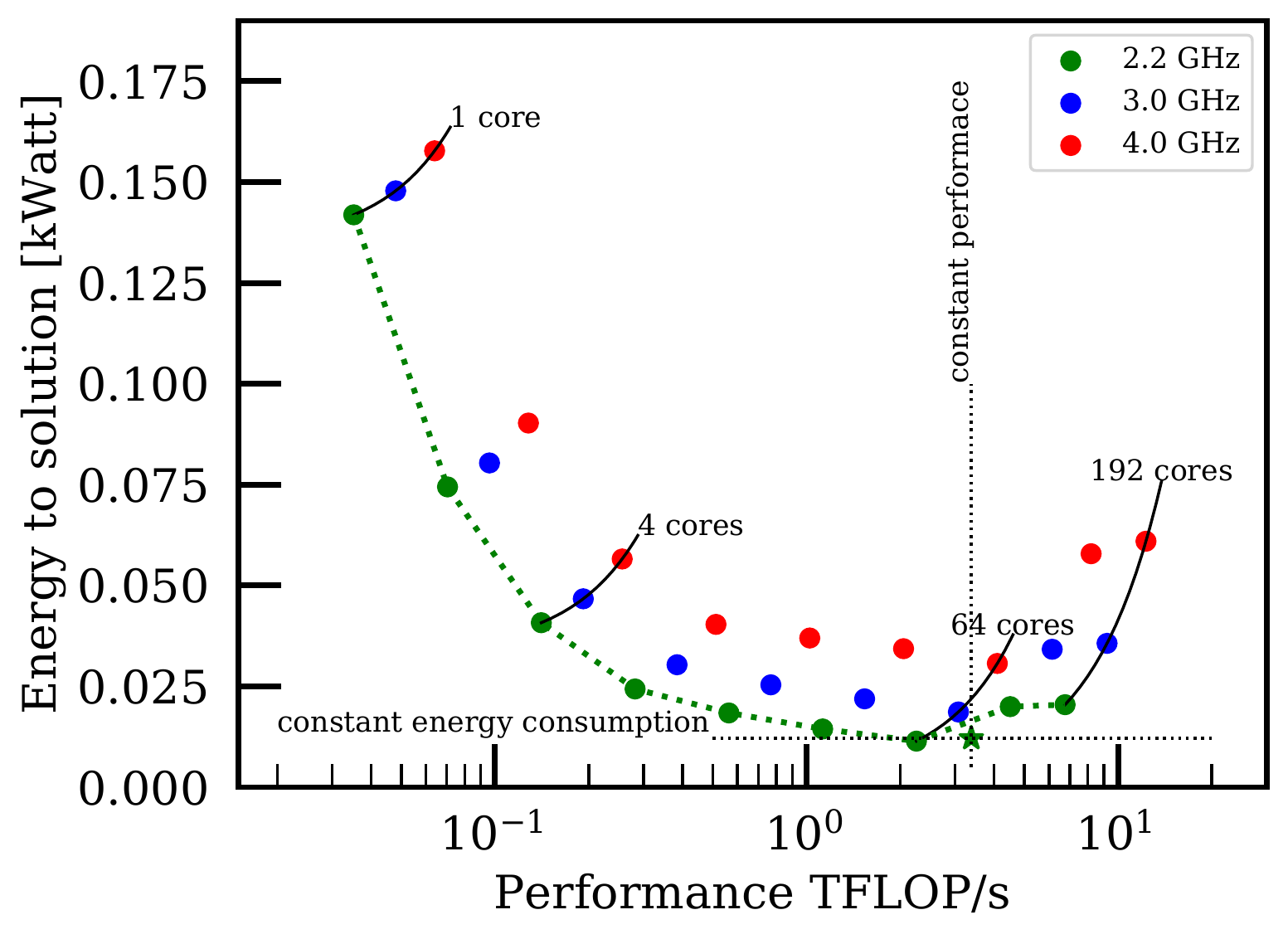}
\caption{\bf Energy to solution as a function of code performance.
  The Z-plot gives for a number of processor (and processor
  frequencies) and the energy consumed (in kWatt) as a function of
  performance (in TFLOP/s) \cite{DBLP:journals/corr/abs-1304-7664}.
  The runs (green dots) were performed using a quad CPU 24-core (48
  hyperthreaded) Intel Xeon E7-8890 v4 at 2.20\,GHz calculated with 1,
  2, 4, ..., 192 cores. Curves of constant core-count are indicated
  for 1, 4, 64 and 192 cores (solid curves).  The other colored points
  (blue and red) give the relation for overclocking the processor to 3
  and 4\,GHz, scaled from the measured points using over-clocking
  emission relations \cite{Overclocking_document}.  Dotted curves
  give constant energy-requirement-to-solution (horizontal) and
  sustained processor performance (vertical). The star at the cross of
  these two curves is measured using 96 cores. The calculations are
  performed Bulirsch-Stoer algorithm with a Leofrog integration
  \cite{2041-8205-785-1-L3} at a tolerance of $\log(dE/E) = -8$
  using a wordlength of 64 bits.}
\label{fig:Zplot}
\end{figure}

When parallelism is optimally utilized, the highest performance is
reached for the maximum core count, but the optimal combination of
performance and carbon emission is reached for $\sim 1000$ cores,
after which the supercomputer starts to produce more carbon than a
workstation. The improved energy characteristics for parallel
operation and its eventual decline is further illustrated in the
Z-plot presented in figure\,\ref{fig:Zplot}, showing energy
consumption as a function of the performance of 96 cores (192
hyperthreaded) workstation.

Running single core on a workstation is inefficiently slow and
produces more carbon than running multi-core. Performance continues to
increase with core count, but optimal energy consumption is reached
when running 64 and 96 physical cores (green star in
figure\,\ref{fig:Zplot}). Running more cores will continue to reduce
the time-to-solution, but at higher emission. Note that the carbon
emission of the parallel calculation (blue curve in
figure\,\ref{fig:tts_CO2_emission}) does not drop, because we assumed
that the supercomputer is optimally shared, whereas we assumed that
the workstation used in the Z-plot figure was private.

When scaling our measurements of the compute performance and energy
consumption with the clock frequency of the processor (blue and red
points for each core-count) reduces wall-clock time, but costs
considerably more energy (see also
\cite{DBLP:journals/corr/abs-1803-01618}). Although not shown here,
reducing clock-speed slows down the computer while increasing the
energy requirement.

If the climate is a concern, prevent occupying a supercomputer to
capacity. The wish for more environmentally friendly supercomputers
triggered the contest for the greenest supercomputers \cite{4404810}.
Since the inauguration of the green500, the performance per Watt has
increased from 0.23\,Tflop/kW by a Blue Gene/L in 2007 \cite{4404810}
to more than 20\,Tflop/kW by the MN-3 core-server today
\cite{Green500}. This enormous increase in performance per Watt is
mediated by the further development of low-power many-core
architectures, such as the GPU. The efficiency of modern
workstations, however, has been lagging. A single core of the
Intel Xeon E7-8890, for example runs at $\sim 4$\,TFLOP/kWatt, and the
popular Intel core-i7 920 tops only 0.43\,TFLOP/kWatt. Workstation
processors have not kept up with the improved carbon characteristics
of GPUs and supercomputers.

For optimal operation, run few ($\sim 1000$) cores on a supercomputer
or a GPU-equipped workstation. When running a workstation, use as many
physical cores as possible, but leave the virtual cores alone.
Over-clocking reduces wall-clock time but at a greater environmental
impact.

\section{The role of language on the ecology}

So far, we assumed that astrophysicists invest in full code
optimization that uses the hardware optimally. However, in practice,
most effort is generally invested in developing the research question,
after which designing, writing, and running the code is not the
primary concern. This holds so long as the code-writing and execution
are sufficiently fast. As a consequence, relatively inefficient
interpreted scripting languages, such as Python, rapidly grow in
popularity.

%% In fact, these percentages refer to the number of bytes in each of
%% these languages.

According to the Astronomical Source Code Library (ASCL \cite{ASCL}),
$\sim 43$\% of the code is written in Python, and 7\,\% Java, IDL and
Mathematica.  Only 18\%, 17\% and 16\% of codes are written in
Fortran, C and C++ respectively. Python is popular because it is
interactive, strongly and dynamically typed, modular, object-oriented,
and portable. But most of all, Python is easy to learn and it gets the
job done without much effort, whereas writing in C++ or Fortran can be
rather elaborate. The expressiveness of Python considerably outranks
the Fortran and C families of programming languages.

\begin{figure}
\includegraphics[width=\columnwidth]{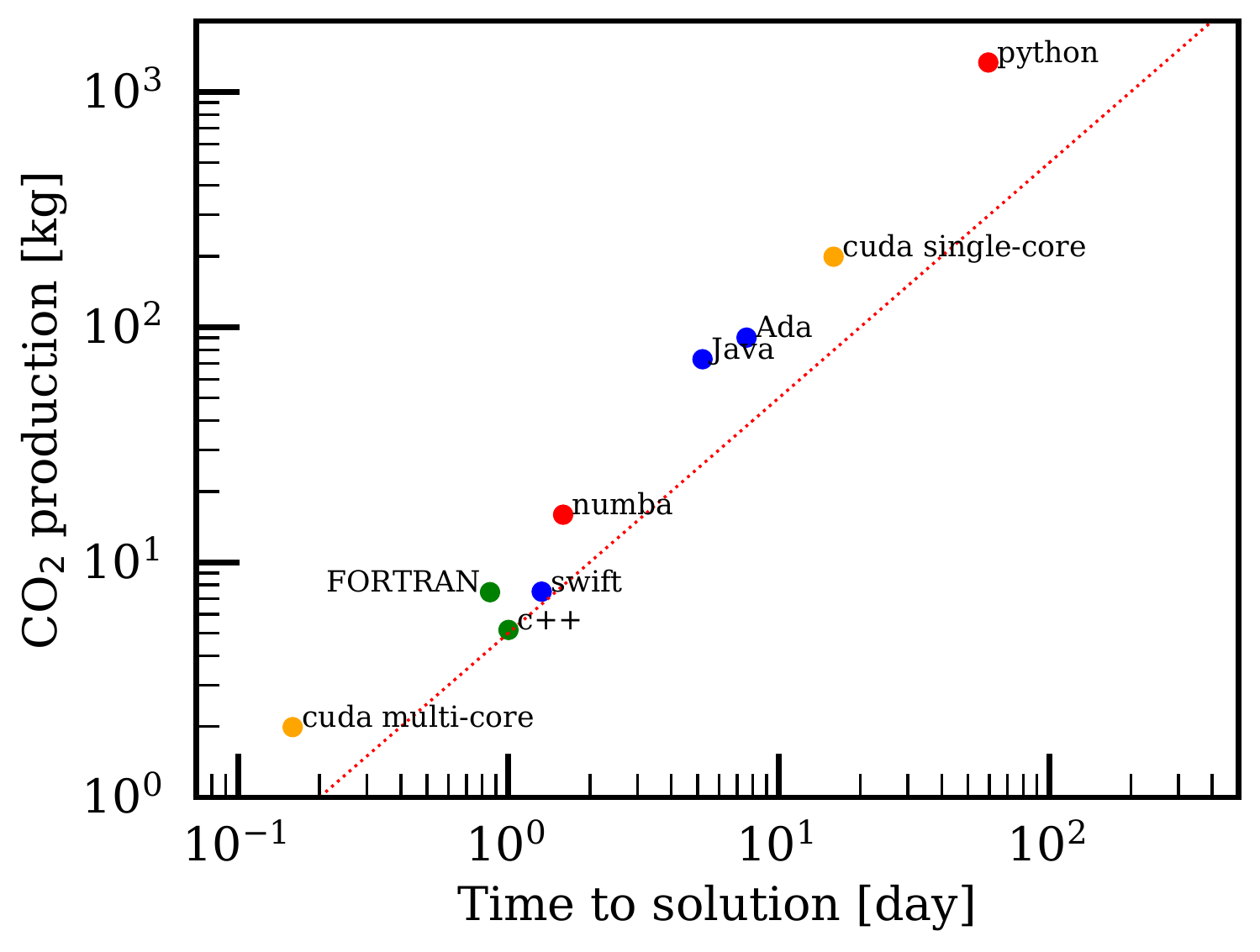}
\caption{\bf Here we used the direct $N$-body code from \cite{Nbabel}
to measure execution speed and the relative energy efficiency for
each programming language from table\,3\, of
\cite{10.1145/3136014.3136031}. The dotted red curve gives a linear
relation between the time-to-solution and carbon footprint ($\sim
5$\,kg\,CO$_2$/day). The calculations were performed on a 2.7GHz
Intel Xeon E-2176M CPU and NVIDIA Tesla P100 GPU. }
\label{fig:Language_efficiency}
\end{figure}

The main disadvantage of Python, however, is its relatively slow speed
compared to compiled languages. In
figure\,\ref{fig:Language_efficiency}, we present an estimate of the
amount of CO$_2$ produced when performing a direct N-body calculation
of $2^{14}$ equal-mass particles in a virialized Plummer sphere. Each
calculation was performed for the same amount of time and scaled to 1
day for the implementation in C++.

Python (and to a lesser extend Java) take considerably more time to
run and produce more CO$_2$ than C++ or Fortran. Python and Java are
also less efficient in terms of energy per operation than compiled
languages \cite{10.1145/3136014.3136031}, which explains the offset
away from the dotted curve.

The growing popularity of Python is disquieting. Among 27 tested
languages, only Perl and Lua are slower
\cite{10.1145/3136014.3136031}. The runtime performance of Python can
be improved in a myriad of ways. Most popular are the numba or NumPy
libraries, which offer pre-compiled code for common operations. In
principle, numba and NumPy can lead to an enormous increase in speed
and reduced carbon emission. However, these libraries are rarely
adopted for reducing carbon emission or runtime with more than an
order of magnitude \cite{ASCL}. NumPy, for example, is mostly used for
its advanced array handling and support functions. Using these will
reduce runtime and, therefore, also carbon emission, but optimization is
generally stopped as soon as the calculation runs within an
unconsciously determined reasonable amount of time, such as the
coffee-refill time-scale or a holiday weekend.

In figure\,\ref{fig:tts_CO2_emission} we presented an estimate of the
carbon emission as a function of runtime for Python implementations
(see blue dotted curve) of popular applications (green and blue
bullets). The continuing popularity of Python should be confronted
with the ecological consequences. We even teach Python to students,
but also researchers accept the performance punch without realizing the
ecological impact. Using C++ and Fortran instead of Python would save
enormously in terms of runtime and CO$_2$ production. Implementing in
CUDA and run on a GPU would even be better for the environment, but
the authors know from first-hand experience that this poses other
challenges, and that it takes years of research
\cite{2007NewA...12..641P}, before a tuned instrument is production-ready \cite{Bedorf:2014:PGT:2683593.2683600}.

\section{Conclusions}

The popularity of computing in research is proliferating. This
impacts the environment by increased carbon emission.

The availability of powerful workstations and running Python scripts
on single cores is about the worst one can do for the
environment. Still, this mode of operation seems to be most popular
among astronomers. This trend is stimulated by the educational system
and mediated by Python's rapid prototyping-abilities and the
ready availability of desktop workstations. This trend leads to an
unnecessarily large carbon footprint for computationally-oriented
astrophysical research. The importance of rapid prototyping
appears to outweigh the ecological impact of inefficient code.

The carbon footprint of computational astrophysics can be reduced
substantially by running on GPUs. The development time of such code,
however, requires major investments in time and requires considerable
expertise.  As an alternative, one could run concurrently using
multiple the cores, rather than a single thread.  It is even better to
port the code to a supercomputer and share the resources. Best
however, for the environment is to abandon Python for a more
environmentally friendly (compiled) programming language. This would
improve runtime and reduces CO$_2$ emission.

There are several excellent alternatives to Python. The first choice
is to utilize high-performance libraries, such as NumPy and Numba. But
there are other interesting strongly-typed languages with
characteristics similar to Python, such as Alice, Julia, Rust, and
Swift. These languages offer the flexibility of Python but with the
performance of compiled C++. Educators may want to reconsider teaching
Python to University students. There are plenty environmentally
friendly alternatives.

While being aware of the ecological impact of high-performance
computing, maybe we should be more reluctant in performing specific
calculations, and consider the environmental consequences before
performing a simulation. What responsibility do scientists have in
assuring that their computing environment is mostly harmless to the
environment?

{\bf Acknoweldgments}

It is a pleasure to thank Alice Allen discussions.

We used the Python \cite{vanRossum:1995:EEP},
matplotlib~\cite{2007CSE.....9...90H},
numpy~\cite{Oliphant2006ANumPy}, and AMUSE~\cite{2018araa.book.....P}
open source packages. Calculations ware performed using the LGM-II
(NWO grant \# 621.016.701), TITAN (LANL), and ALICE (Leiden
University).

%%\input /home/spz/Latex/lib/bib/references

\end{document}